\documentclass[aip,amsmath,amssymb,floatfix,
preprint,%
]{revtex4-1}

\usepackage{graphicx}
\usepackage{dcolumn}
\usepackage{bm}
\usepackage[utf8]{inputenc}
\usepackage[T1]{fontenc}
\usepackage{mathptmx}
\usepackage{subcaption}
\usepackage{etoolbox}
\usepackage{siunitx}
\usepackage{tcolorbox}
\usepackage{tikz}
\usepackage{xcolor}
\usepackage{float}

\DeclareSIUnit{\atm}{\mathrm{atm}}

\definecolor{NewRed}{HTML}{C60C1F}

\begin{document}

\preprint{AIP/123-QED}

\title[Shock-induced cavitation in a water droplet]{Shock-induced cavitation and wavefront analysis inside a water droplet}

\author{Luc Biasiori-Poulanges}%
\email{lbiasiori@ethz.ch}%
\affiliation{Institute of Fluid Dynamics, Department of Mechanical and Process Engineering, ETH Zurich\\ Sonneggstrasse 3, 8092 Z\"urich, Switzerland}%
\author{Hazem El-Rabii}%
\email{hazem.elrabii@cnrs.pprime.fr}%
\affiliation{Institut Pprime, CNRS UPR 3346 -- Universit\'e de Poitiers -- ISAE-ENSMA,\\ 1 avenue Cl\'ement Ader, 86961 Futuroscope, France}%

\date{\today}

\begin{abstract}
The objective of the present study is to develop a basic understanding of the interaction of shock waves with density inhomogeneities. We consider here the particular instance of a planar air shock impinging on a spherical water-droplet and discuss to what extent this interaction can lead to the inception of cavitation inside the droplet. The effort centers on the early phases of the interaction process during which the geometry and amplitude of the propagating wavefront is modified by refraction and the subsequent internal reflections at the droplet interface. The problem is analysed using both simple ray theory and a 2-D multiphase, compressible hydrodynamic code (ECOGEN). Within the context of ray theory, the occurrence of focusing is examined in details and parametric equations are derived for the transmitted wavefront and its multiple internal reflections. It is found that wave patterns predicted by ray calculations compare extremely well with the more accurate numerical solutions from simulations. In particular, it is shown that the internal wavefront assumes a complex time-dependent shape whose dominant feature is the existence of cusp singularities. These singular points are shown to trace out surfaces that are the caustics of the associated system of rays. From the singularities of the energy flux density of the refracted wave, the parametric equations of the caustic surface associated to the $k$-th reflected wavefront are deduced. As a consequence of the focusing process, the simulations show the formation of negative-pressure regions in the internal flow field. These low-pressure zones are identified as possible spots where cavitation may occur, depending on the magnitude of the pressure reached. Finally, the numerical results provide quantitative information on the dependence of negative pressure peak upon incident-shock-wave strength.
\end{abstract}

\maketitle

\section{Introduction}
\label{sec:intro}
The fundamental mechanisms governing aerobreakup have been addressed in numerous studies where the fragmentation of a single spherical drop suddenly exposed to a uniform high-speed gas flow was considered\cite{guildenbecher2009secondary}. The relative velocity of the drop with respect to the ambient flow field has often been realized by its injection into the uniform flow field behind a shock wave. Curiously enough, the question whether the shock wave itself may have any effect on the deformation and breakup process has not received much attention. The underlying reasons likely are twofold: first, the time it takes for the shock wave to transit the drop is too short to cause any significant drop response during the interaction\cite{aalburg_deformation_2003}. Secondly, the large difference in the shock impedance between the ambient gas and the liquid results in a poor energy transfer into the liquid (e.g., transmission coefficient from air to water $\approx0.1\%$). While it is tempting to conclude from the above that the shock wave has no direct effect on the droplet evolution, a closer consideration of the matter shows that the answer is not that straightforward.

Indeed, because of the large shock impedance contrast between air and water, the interface bounding the liquid medium acts as a perfect mirror trapping the transmitted wave energy within the droplet. As a result, the confined shock wave experiences nearly-total reflections and focusing that amplifies its local interaction with the liquid on short time scales. The important point here is that the reflected wave is a focused expansion wave that can, under some conditions, expose regions of the liquid to a pulling force. This suggests the possibility for the liquid to cavitate. Water, for instance, cannot withstand significant tension and starts to cavitate whenever pressure falls below some critical value. Given that the presence of vapour cavities inside liquid droplets alters the interfacial dynamics\cite{liang_interaction_2020,biasiori2020multimodal}, changes in the fragmentation process are to be expected, especially if high-speed jets develop during cavity collapsing\cite{liang_interaction_2020}.

In this paper, the question we are concerned with is under what conditions low-enough negative pressure to cause cavitation can be reached inside a shock impacted water droplet. We use numerical simulations to identify these conditions by considering that cavitation starts whenever pressure falls below some critical value. The simulations' results and, in particular, the complex wavefront patterns generated inside the droplet are interpreted qualitatively using the classical ray-tracing approach to geometrical acoustics.

\section{Confined wavefront propagation}
We consider a planar shock wave propagating through air and impinging on a spherical water droplet. The shock-droplet interaction results in a shock that is transmitted through the interior of the water droplet while a portion of the incident shock diffracts around the edge of the droplet. Here, we focus on the transmitted shock and analyse the wavefront evolution following a dynamic ray-tracing method \citep{cerveny2005seismic}. This approach consists in studying shock propagation on the basis of the concept of rays considered as orthogonal trajectories along which wavefronts travel. Both media are assumed homogeneous so that rays are straight lines along which wavefronts propagate at constant speed. 
\begin{figure}[h]
	\centering
	\includegraphics[width=250pt]{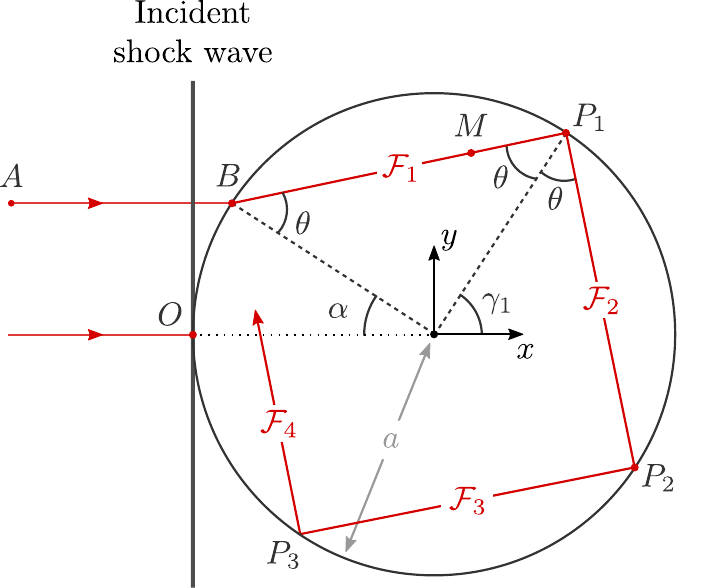}
	\caption{Ray diagram showing the refraction and the multiple internal reflection of initially parallel rays incident crossing the droplet boundary.}
	\label{fig:angle}
\end{figure}

In accordance to the ray formalism, the incident shock wave propagates along a family of parallel rays incident from the right onto the droplet. Figure~\ref{fig:angle} illustrates the geometry of the problem. The radius of the spherical droplet is denoted by $a$ and its center is at the origin of the coordinate system. We conveniently choose the $x$-axis in the direction of the incoming parallel bundle of rays and the time origin as the instant at which the shock wave reaches $x=-a$. Consider now an arbitrary ray $AB$ striking the droplet surface at point $B$ with an incident angle $\alpha$. The major part of the ray amplitude is reflected at $B$, while the remaining part of the intensity is transmitted into the droplet. The refracted ray makes an angle $\theta$ with the interface normal at $B$. The incident and refracted angles are related by the fundamental law of refraction \citep{henderson_refraction_1989}: 
\begin{equation}\label{eq:snell}
\sin\theta=n\,\sin\alpha.
\end{equation}
Here, $n$ is the ratio of wave velocity in water to that of air. The transmitted ray is then internally reflected at each interaction with the droplet surface at angle to normal of $\theta$. The typical ray path of the refracted wavefront consists therefore of many successive segments separated by reflection points, $P_{k}$, at the droplet boundary. Following usage in geometrical optics, we define a $k$-ray family to be  rays within the droplet (for all $\alpha$) that have undergone  $k-1$ internal reflections \cite{adler1997high}. According to this definition, the transmitted rays that have not suffered internal reflection belong to the 1-ray family. They become the 2-rays after their first reflection, and so on. The wavefront travelling along a $k$-ray family will be denoted by $\mathcal{F}_{k}$.

We shall now derive the parametric equations for $\mathcal{F}_{k}$. To that end, let us consider an arbitrary point $M$ belonging to a $k$-ray. The location of this point is expressed by its position vector as
\begin{equation}\label{eq:vectorposM}
	\mathbf{r}_M = \mathbf{r}_{P_{k-1}} + \mathbf{e}_{k}\,\ell_{P_{k-1}M},
\end{equation}
where $\mathbf{e}_{k}$ are the unit vectors in the $k$-ray direction, and $\ell_{P_{k-1}M}$ is the length of the segment $P_{k-1}M$. Elementary geometrical considerations (Fig.~\ref{fig:angle}) yield the coordinates of the $k$-th internal reflection point, $P_k$, 
\begin{equation}\label{eq:pk}
	(x_{P_k},y_{P_k}) =  (a \cos\gamma_k,\,a \sin\gamma_k),
\end{equation}
where $\gamma_k = 2k\theta-\alpha-(k-1)\pi$. $P_0$ corresponds to the entry point ($B$) of ray $AB$ into the droplet. Using (\ref{eq:pk}) we can then write down at once that
\begin{equation}
	\mathbf{e}_{k} = \mathbf{e}_{x}\, \cos(\gamma_k-\theta) +\mathbf{e}_{y}\, \sin(\gamma_k-\theta),\quad k \ge 1,
\end{equation}
where $\mathbf{e}_{x}$ and $\mathbf{e}_{y}$ are the unit vectors in $x$- and $y$-direction, respectively.
\begin{figure*}[tbph]
	\centering
	\includegraphics[width=\textwidth]{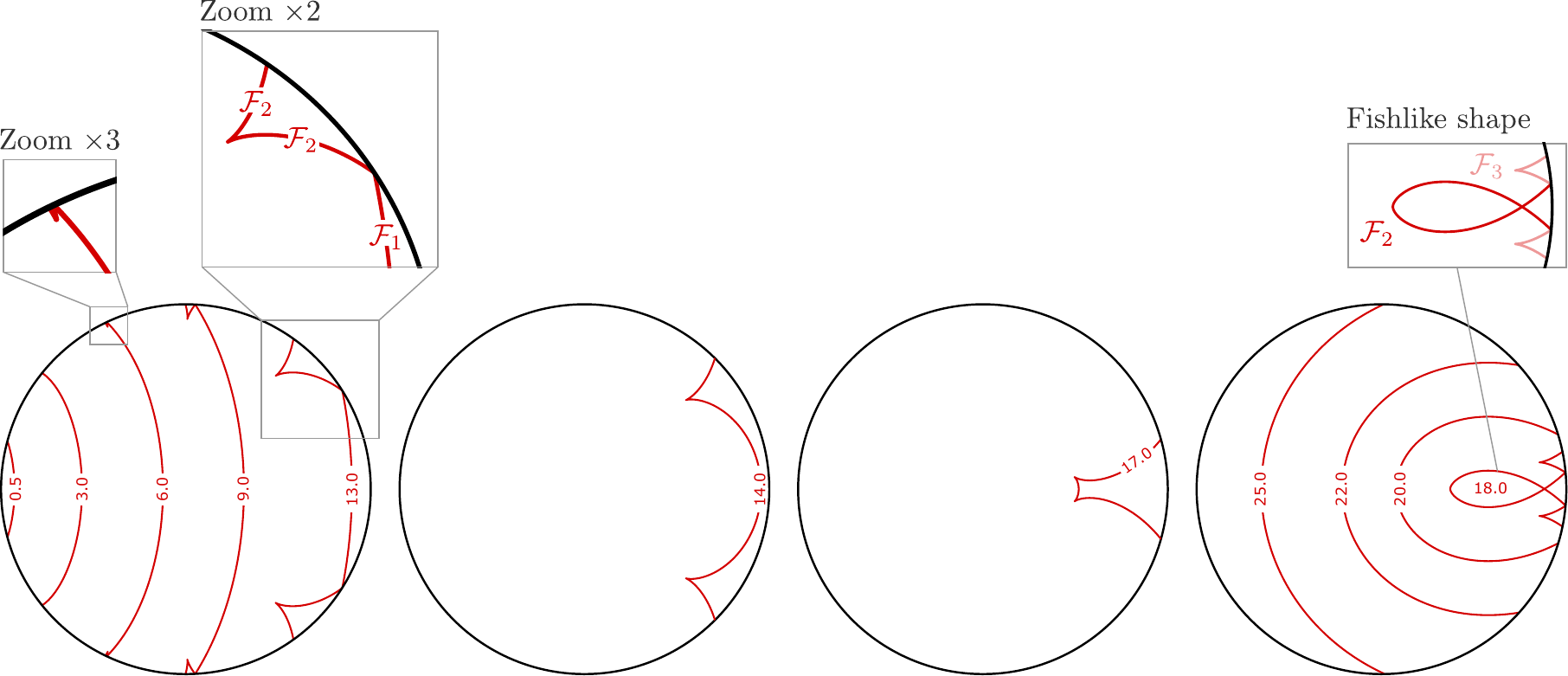}
	\caption{Propagation of the confined wavefront within the droplet ($n=2.12$). 	Labels on contour indicates the physical time in tenth of microsecond. Wavefronts $\mathcal{F}_{k=1,2}$ are plotted from parametric equations~(\ref{eq:refracted_front}). The case $k=3$ only shown for $t=\SI{1.8}{\micro\second}$ is intended to be illustrative.}
	\label{fig:AcousticGeom}
\end{figure*}

To determine $\ell_{P_{k-1}M}$, we note that the time $t$ required for point $M$ on the $k$-ray to be reached by the wavefront $\mathcal{F}_k$ is
\begin{equation}\label{eq:timeM}
t = \frac{a}{u_\mathrm{a}} (1-\cos\alpha) + 2 (k-1) \frac{a}{u_\mathrm{w}}\cos\theta + \frac{\ell_{P_{k-1}M}}{u_\mathrm{w}}.
\end{equation}
The first term in (\ref{eq:timeM}) is the time for the front to travel along the ray $AB$ from $x=-a$ to $B$, whilst the second term represents the total travel time it takes for the front to go from $B$ to $P_{k-1}$ along the segment-rays in between. Consequently, Eq.~(\ref{eq:timeM}) yields
\begin{equation}\label{eq:lpm}
\ell_{P_{k-1}M} = u_\mathrm{w}\,t - n\, a \,(1-\cos\alpha) - 2 (k-1) a \cos\theta.
\end{equation}

After substituting (\ref{eq:lpm}) into (\ref{eq:vectorposM}), and expressing the result in Cartesian components we find 
\begin{subequations}\label{eq:refracted_front}
	\begin{eqnarray}
	x_M & = &  (u_\mathrm{w}\,t - n\,a\,(1-\cos\alpha) - 2 (k-1) a \cos\theta )\,\cos(\gamma_k-\theta)-a \cos(\gamma_{k}-2\theta),\label{eq:refracted_front_x}\\
	y_M & = & (u_\mathrm{w}\,t - n\,a\,(1-\cos\alpha) - 2 (k-1) a \cos\theta )\,\,\sin(\gamma_k-\theta)-\,\,a \sin(\gamma_{k}-2\theta).\label{eq:refracted_front_y}
	\end{eqnarray}
\end{subequations}

Given that the wavefront shape is defined as the locus of points reached by a disturbance in a given time along all possible ray paths, Eqs~(\ref{eq:refracted_front_x})--(\ref{eq:refracted_front_y}) represent the parametric equations of the wavefront $\mathcal{F}_k$, with $\alpha$ as parameter.

Since our interest is the determination of the refracted wavefront and its internal reflections, we need consider only the incident rays that meet the upstream droplet surface at an angle lower than the critical angle for total reflection, that is, $|\alpha| < \alpha_c=\arcsin(1/n)$. It is to be noted that the range of $\alpha$ values is not restricted solely by $\alpha_c$. Indeed, for a $k$-ray family, two specific rays ($\mathcal{R}_l$ and $\mathcal{R}_u$) bound the region in which the $k$-rays lie. The ray $\mathcal{R}_l$ corresponds to the ray reaching the point $P_{k-1}$ at time $t$, whereas $\mathcal{R}_u$ is the ray that hits, at the same instant, the inner droplet surface at $P_{k}$. The associated incident angles, $\alpha_l$ and $\alpha_u$, are obtained by setting $\ell_{P_{k-1}M}$ equals to 0 and $2 a \cos\theta$, respectively, into (\ref{eq:lpm}):
\begin{eqnarray}
	2\, (k-1) a \cos\theta_l & = & u_\mathrm{w}\,t - n\,a\,(1-\cos\alpha_l),\label{eq:alphal}\\
	2\, k a \cos\theta_u & = & u_\mathrm{w}\,t - n\,a\,(1-\cos\alpha_u),\label{eq:alphau}
\end{eqnarray}
where $\theta_l$ and $\theta_u$ are the refraction angles corresponding to $\alpha_l$ and $\alpha_u$, respectively. Equations (\ref{eq:alphal}) and (\ref{eq:alphau}) can be solved exactly, as they are quadratic in $\cos\alpha$. The values of $\alpha$ that should be considered for $k=1$ are such that $|\alpha|$ is bounded by the lowest value between $\alpha_c$, $\alpha_l$, and $\alpha_u$. For $k$ greater than one, the absolute value of $\alpha$ is between $\alpha_l$ and $\alpha_u$.

Figure~\ref{fig:AcousticGeom} shows the wavefront pattern generated inside the droplet, at different instants, as calculated from Eqs.~(\ref{eq:refracted_front_x})--(\ref{eq:refracted_front_y}) for $k=1$ and 2. To avoid overloading the figure with crossed and/or juxtaposed fronts, the successive wavefront positions are displayed in different panels: (a) $\SI{50}{\nano\second} \le t \le \SI{1.3}{\micro\second}$, (b) $t=\SI{1.5}{\micro\second}$, (c) $t=\SI{1.7}{\micro\second}$, (d) $\SI{1.8}{\micro\second} \le t \le \SI{2.5}{\micro\second}$. We observe from Fig.~\ref{fig:AcousticGeom}(a) that the transmitted front appears as originating from an external point source located on the symmetry axis. It does not exhibit any singular point during early times ($t\lesssim\SI{0.6}{\micro\second}$), i.e., the front shape is smooth. Closer examination of the front shape at $t=\SI{0.6}{\micro\second}$  reveals that the wavefront folds itself where it is in contact with the droplet boundary. The fold moves along the boundary as $\mathcal{F}$ propagates and splits $\mathcal{F}$ into two sub-fronts, $\mathcal{F}_1$ and $\mathcal{F}_2$. The segment of the front ahead of the fold ($\mathcal{F}_1$) remains smooth all along its propagation and corresponds to the rays that have experienced only a single refraction. The front segment $\mathcal{F}_2$ starts to develop simultaneously with the appearence of the fold, near the droplet surface (see Fig.~\ref{fig:AcousticGeom}(a), $t=\SI{0.6}{\micro\second}$). Contrary to $\mathcal{F}_1$, $\mathcal{F}_2$ exhibits a singular point (cusp) as it is particularly apparent at $t\simeq\SI{0.9}{\micro\second}$ and grows as the front travels through the droplet.
\begin{figure*}[h]
    \centering
    \includegraphics[width=.8\textwidth]{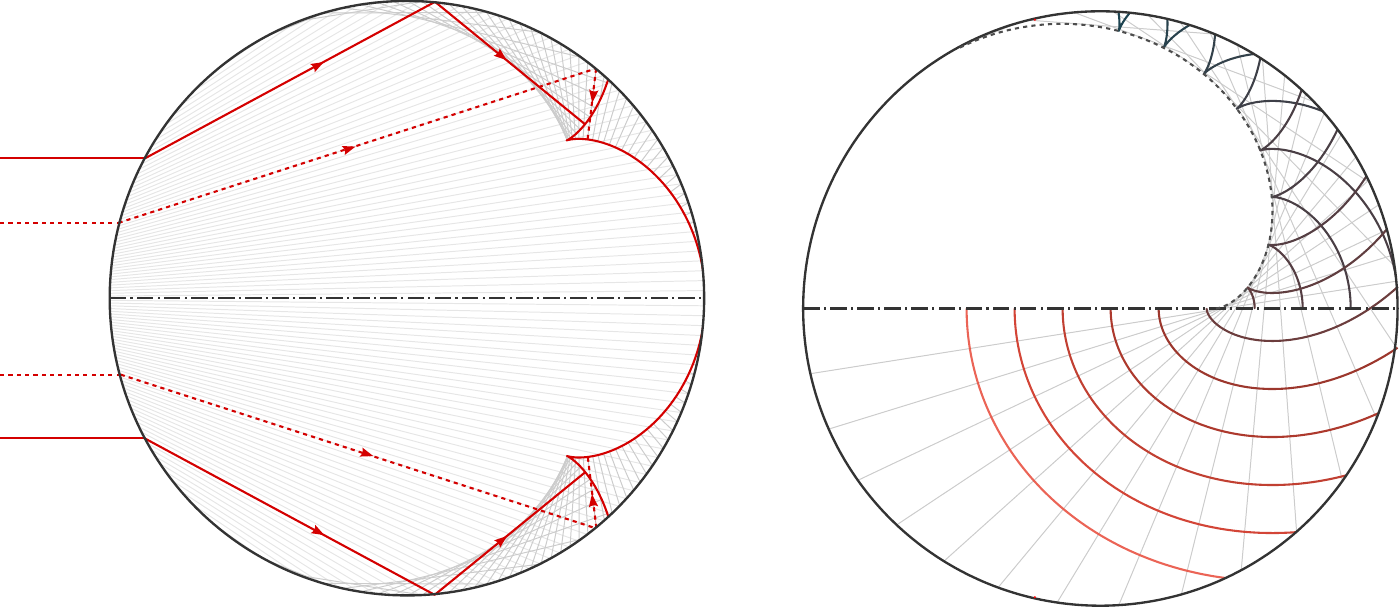}
    \caption{(Left) Illustration of the $k$-ray family (gray) and the corresponding $\mathcal{F}_{k=1,2}$ wavefronts (red). Four rays are highlighted in red solid and dashed lines for readability purposes. (Right) Superposition of successive fronts $\mathcal{F}_2$ that evidences the cusp motion and the caustic traced out (black dashed line). The black-to-red color scale is representative of the physical time.}
    \label{fig:F1_time}
\end{figure*}
On reaching the downstream droplet surface, $\mathcal{F}$ is completely reflected back. The fronts displayed in Fig.~\ref{fig:AcousticGeom}(b)--(d) are thus exclusively once-reflected fronts (i.e., $\mathcal{F}_2$), which travel from right to left. As time proceeds, we see from Fig.~\ref{fig:AcousticGeom}(b) and (c) that the cusps from either side of the symmetry axis get closer to each other. The front $\mathcal{F}$ then passes through itself, developing a self-intersecting swallowtail pattern (not shown here). This cusp motion is accompanied by a focusing of the front segment connecting the pair of cusps, until we observe cusp annihilation. Subsequently, $\mathcal{F}$ takes on a fishlike shape before it begins to diverge and becomes smooth again, Fig.~\ref{fig:AcousticGeom}(d).
A striking feature of the cusp motion is evidenced by superposing successive fronts $\mathcal{F}_2$, as shown in Fig.~\ref{fig:F1_time}. Indeed, as the wavefront advances, we see that the cusp of $\mathcal{F}_2$ traces out a curve, the so-called caustic, which is shown by a dashed line in Fig.~\ref{fig:F1_time}. We notice that the caustic has a cusp singularity at $C$, where it is clearly seen that the front's cusp cancels.

\section{Caustics inside the liquid droplet}

It is well known that a caustic corresponds to regions where several rays bunch together to form discontinuities at which the intensity diverges. This means that the front's cusp is a moving focus, and the caustic is the surface traced by it. 

\begin{figure}
	\centering
    \includegraphics[scale=0.7]{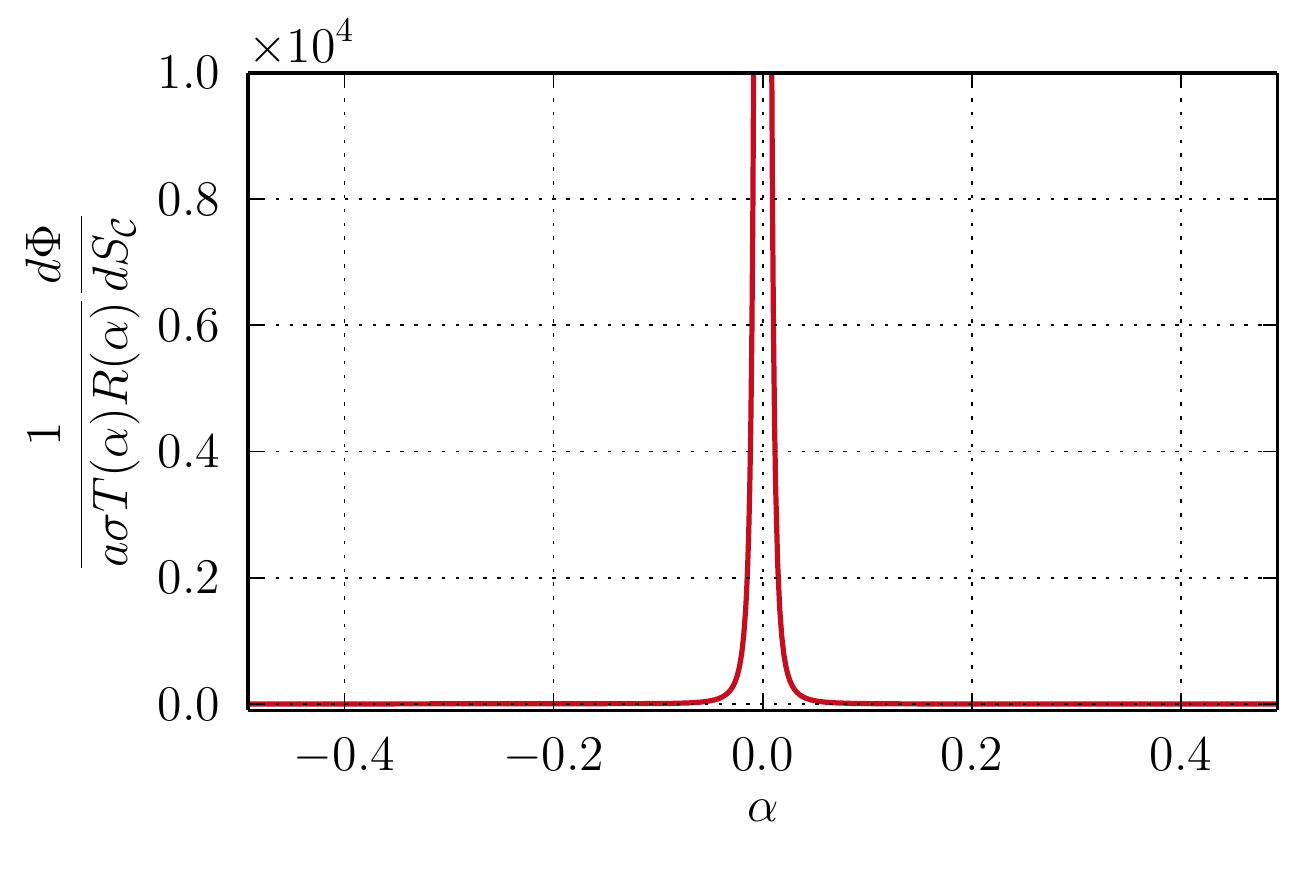}
 	\caption{Variation of the density of rays tangent to the caustic (for $k=2$ and $n=2.12$) over the incident angle $\alpha$. The concentration increases when $\alpha$ decreases, becoming infinite for $\alpha=0$.}
 	\label{fig:energy-density}
\end{figure}

To determine the parametric equation of the caustic, we note that the loci of points of high ray concentration can be obtained from the singularities of the energy flux density of the transmitted wave. If $\sigma$ denotes the flux density associated with the plane incident wave, the flux incident upon an element of area $dS_\mathrm{d}$ on the droplet surface is $d\Phi = \sigma \cos\alpha\,dS_\mathrm{d}$. Due to the axial symmetry about the $x$-axis, we have that $dS_\mathrm{d}=2\pi a^2 \sin\alpha\,d\alpha$. The fraction of $d\Phi$ that survives $k-1$ internal reflections is thus given by
\begin{eqnarray}
	d\Phi_M & = & T(\alpha)\,R^{k-1}(\alpha)\, d\Phi\nonumber\\
	        & = & 2\pi a^2 \sigma\,T(\alpha)\,R^{k-1}(\alpha)\, \sin\alpha \cos\alpha\,d\alpha,
\end{eqnarray}
where $T(\alpha)$ and $R(\alpha)$ are the transmission and reflection coefficient. The flux density over the transmitted wavefront is equal to $d\Phi_M$ divided by the element of area $dS$ mapped out by the rays that have crossed $dS_\mathrm{d}$:
\begin{equation}\label{eq:fluxdensity}
	\frac{d\Phi_M}{dS} = \frac{a^2 \sigma\,T(\alpha)\,R^{k-1}(\alpha)\, \sin\alpha \cos\alpha}{y_M (\dot{x}_M^2+\dot{y}_M^2)^{1/2}}.
\end{equation}
The superposed dot indicates derivative with respect to $\alpha$. Inserting Eqs.(\ref{eq:refracted_front_x})--(\ref{eq:refracted_front_y}) in (\ref{eq:fluxdensity}), we find that
\begin{equation}\label{eq:fluxdensitybis}
	\frac{d\Phi_M}{dS} = \frac{a^2 \sigma\,T(\alpha)\,R^{k-1}(\alpha)\, \sin\alpha \cos\alpha}{(\ell_{P_{k-1}M} \,\sin(\gamma_k-\theta)+ a \sin\gamma_{k-1})\, \left|(\dot{\gamma}_{k}-\dot{\theta})\ell_{P_{k-1}M}-a \dot{\gamma}_{k-1}\cos\theta\right| }.
\end{equation}
The parametric equations for the caustic surfaces are obtained from the condition that the denominator of expression (\ref{eq:fluxdensitybis}) be zero. Thus,
\begin{eqnarray}
	(\dot{\gamma}_{k}-\dot{\theta})\ell_{P_{k-1}M}-a \dot{\gamma}_{k-1}\cos\theta & = & 0,\label{eq:cuspcaustic}\\
	\ell_{P_{k-1}M} \,\sin(\gamma_k-\theta)+a \sin\gamma_{k-1} & = & 0. \label{eq:axialcaustic}
\end{eqnarray}
The condition (\ref{eq:cuspcaustic}) relates the angle value corresponding to the front cusp at time $t$, which after some algebra can be recast as
\begin{equation}\label{eq:condcuspcaustic}
    2 n\, \sin^2\left(\frac{\alpha}{2}\right)+\,2 (f(\alpha)+k-1)\, \cos\theta = \frac{u_\mathrm{w}\, t}{a},{\tiny }
\end{equation}
where 
\begin{equation*}
2 f(\alpha) =  \frac{2 n^2 (k-1) \sin{2\alpha}-\sin{2\theta}}{n^2 (2k-1) \sin{2\alpha}-\sin{2\theta}}.
\end{equation*}
If we eliminate $t$ from Eqs. (\ref{eq:refracted_front_x})--(\ref{eq:refracted_front_y}) by means of the condition (\ref{eq:condcuspcaustic}), we obtain for the caustic of order $k$
\begin{subequations}\label{eq:caustic}
    \begin{eqnarray}
	    x_\mathcal{C} & = & a f(\alpha)\,\cos\gamma_k + a (f(\alpha)-1) \,\cos(\gamma_k-2\theta),\label{eq:xcaustic}\\
	    y_\mathcal{C} & = & a f(\alpha)\,\,\,\sin\gamma_k + a (f(\alpha)-1) \,\,\sin(\gamma_k-2\theta).\label{eq:ycaustic}
    \end{eqnarray}
\end{subequations}
Since the flux density (\ref{eq:fluxdensitybis}) becomes infinite at the caustic $\mathcal{C}$ it cannot be used to quantify the density of rays at the caustic. Following \citet{burkhard1982}, we therefore compute the density of rays tangent to the caustic, which gives a relative measure of the focusing strength over the caustic. This quantity is obtained by dividing an element of incident flux by the area of the caustic formed by the associated rays, $dS_\mathcal{C}$. For the caustic of singly-reflected rays, we find
\begin{equation}\label{eq:fluxdensitycaustic}
\frac{d\Phi}{dS_\mathcal{C}} = \frac{2a \sigma\,T(\alpha)\,R(\alpha)\, \cos\theta\,\, (\cos\theta-3 n \cos\alpha)^2}
{3 n^2 (6 + 5 n^2 + 11 n^2 \cos 2\alpha - 
	18 n \cos\alpha \cos\theta) \left|y_\mathcal{C}\right|}.
\end{equation}
It is apparent from this expression that the concentration of tangent rays is the highest at the intersection of the caustic and the symmetry axis ($y_\mathcal{C} =0$), that is, at the caustic's cusp. This is because the degree of focusing at the caustic's cusp is higher than over a small element of the caustic's surface area. As exemplified in Fig.~\ref{fig:energy-density} (for $k=2$), the concentration increases when $\alpha$ decreases, becoming infinite for $\alpha=0$. The location of the cuspidal point of the caustic can readily be found by setting $\alpha=0$ in (\ref{eq:xcaustic}), which gives for the horizontal coordinate
\begin{equation}\label{eq:xcusp}
x_\mathrm{cusp} = \frac{(-1)^k \, n}{(2k-1) n-1}\,a\stackrel{k=2}{=}\frac{n}{3 n-1}\,a.
\end{equation}

Equation (\ref{eq:xcusp}) shows that the sign of the abscissa $x_\mathrm{cusp}$ is determined by the parity of $k$, and the position of caustic's cusp gradually approaches the point $O$ with increasing $k$.

It is also interesting to consider the time, $t_f$, at which focusing at this point occurs. As shown in Fig.~\ref{fig:F1_time}(b), $t_f$ coincides with the instant when the two cusps of the once-reflected front merge. This time can be determined from the condition (\ref{eq:condcuspcaustic}). As the left-hand side of the latter is symmetric in $\alpha$ with a maximum for $\alpha$ equals to zero, the sought time is obtained by evaluating (\ref{eq:condcuspcaustic}) at $\alpha=0$, yielding
\begin{equation}\label{eq:tcusp}
t_f = \frac{4k(k-1)\,n + 1-2k}{(2k-1)\,n-1}~\frac{a}{u_\mathrm{w}}\stackrel{k=2}{=}\frac{8 n-3}{3 n-1}~\frac{a}{u_\mathrm{w}}.
\end{equation}

Beside the cusp trajectory described by Eqs.~(\ref{eq:xcaustic})--(\ref{eq:ycaustic}), there exists another region where the flux density (\ref{eq:fluxdensitybis}) is singular. The equations of the loci of these singularities can be determined by using (\ref{eq:axialcaustic}) to eliminate the time $t$ from Eqs.~(\ref{eq:refracted_front_x}).--(\ref{eq:refracted_front_y}). In doing so, we find
\begin{equation}\label{eq:axialcausticbis}
	x_\mathcal{A} = -\frac{a \sin\theta}{\sin(\gamma_k-\theta)},\qquad y_\mathcal{A} = 0,
\end{equation}
which are parametric equations of a straight line segment along the $x$-axis. This high density region results from the focusing on the $x$-axis of incident rays that enter the droplet over a ring of constant $\alpha$. It is straightforward to see that the highest degree of focusing in this case is also achieved at the point of coordinates $(x_\mathrm{cusp},0)$ and time $t_f$.

The determination of the pressure amplitude on caustic surfaces will be described in the next section.

\section{Cavitation inside a water droplet}

As we have seen in the preceding section, ray theory provides a direct physical interpretation of the wave patterns observed within the droplet. Furthermore, it enables to determine regions of the pressure field where the wave is focused (i.e. caustics). The quantitative prediction of pressures at caustics is, however, beyond the scope of ray calculations, which indicate infinite pressure in caustic regions, see Eq.~(\ref{eq:fluxdensitybis}). To identify conditions inducing cavitation inside a droplet hitted by a shock wave, it is then necessary to complement the ray approach with numerical simulations to determine the pressure on caustic surfaces. Before proceeding, however, it is useful to say a few words on the cavitation threshold. 

It is well-known that liquids rupture (or cavitate) when subjected to tensions in excess to some critical tensile that depends on the nature of the liquid and its purity. For pure liquids, cavitation arises from microscopic voids caused by random thermal motions of the molecules.\cite{balibar2002} The process of vapor bubble formation by this mechanism is referred in the literature to as homogeneous nucleation. In contrast, when liquids contain impurities, the maximum tensile they can withstand drastically decreases. This process, termed heterogeneous nucleation, results from the expansion of submicroscopic gas pockets trapped at the solid/liquid interface on the wall of the container or on particles present in the liquid. Water, in particular, has a wide range of measured tensile limits. The maximum tensile that pure water can withstand is \SI{134}{\mega\pascal} at \SI{300}{\kelvin}, according to vapor nucleation theory.\cite{fisher1948} Such a high tensile value has been achieved experimentally.\cite{zheng1991} For not-purified water, the tensile limit becomes less specific and is found to be a few orders of magnitude lower, 0.1--\SI{1}{\mega\pascal}.\cite{caupin_cavitation_2006} Given such a disparity in tensile limits for water, a pressure cavitation threshold has to be chosen, somewhat arbitrarily, within the range of data reported in the scientific literature. We have opted to consider two different values. As far as pure water is concerned, a natural choice is the above-mentioned theoretical limit of \SI{-134}{\mega\pascal}, which will be denoted by $p_\mathrm{c, 1}$. To address the case of not-purified water it is also necessary to consider a higher value as a threshold, $p_\mathrm{c, 2}$. We set this value to \SI{-2.3}{\mega\pascal} on the basis of the experimental results reported by \citet{sembian_plane_2016}

We simulate the  interaction of a planar air shock wave with a spherical water drop using the open-source hydrodynamics code ECOGEN\cite{schmidmayer2020ecogen}. In this code, the dynamics of water and air are modelled using compressible multicomponent flows in which fluid components are assumed immiscible\cite{saurel2009simple}. The water obeys the stiffened gas equation of state with the parameters given in Ref.~\cite{dorschner_formation_2020}, whereas air follows the ideal gas law. Viscous and capillary effects are accounted for according to \citet{schmidmayer2017model}, while phase changes are not modelled. An interface-capturing scheme is used, combining the flow model with a shock-capturing finite-volume method. Additionally, three levels of refinement were used in order to resolve the flow discontinuities. We refer the reader to Refs.~\cite{schmidmayer2020ecogen,dorschner_formation_2020}, and references therein for more details.
\begin{figure}[H]
	\centering
	\includegraphics[width=300pt]{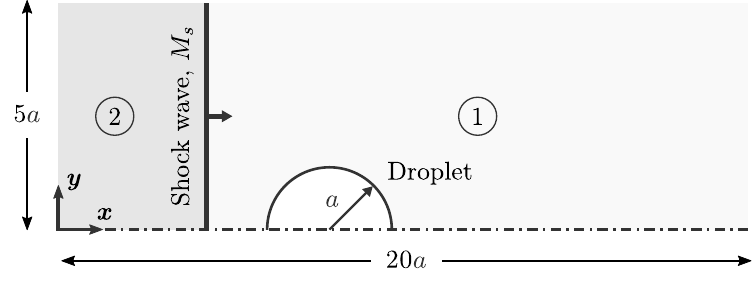}
	\caption{2-D axisymmetric computational domain setup. States 1 and 2 refer to the pre-shock and post-shock conditions, respectively.}
	\label{fig:NumSetUp}
\end{figure}

The problem at hand is treated with a two-dimensional, axisymmetric formulation. A schematic diagram of the computational domain is illustrated in Fig.~\ref{fig:NumSetUp}, where the $x$-axis is the symmetry axis on which the center of the spherical droplet of radius $a$ is located. A symmetric boundary condition is applied to the bottom side of the computational domain and non-reflective boundary conditions are used for the remaining boundaries to avoid contamination of calculations from the reflected outgoing waves. \cite{thompson1987time,thompson1990time,meng_numerical_2015} The shock is initialized inside the domain and travels from left to right. For a given incident shock Mach number $M_s$, the initial flow field is determined from the Rankine-Hugoniot jump relations for an ideal gas using a downstream density of \SI{1.204}{\kilogram\per\cubic\meter}, pressure of \SI{1}{\atm}, and water density of \SI{e3}{\kilogram\per\cubic\meter}. The surface tension between air and water is \SI{72}{\milli\newton\per\meter}. The water drop is assumed to be in mechanical equilibrium with the surrounding air. The excess of pressure inside the droplet over the ambient pressure was computed by employing the Laplace-Young equation. 

\begin{figure*}[htp]
	\centering
	\includegraphics[width=\textwidth]{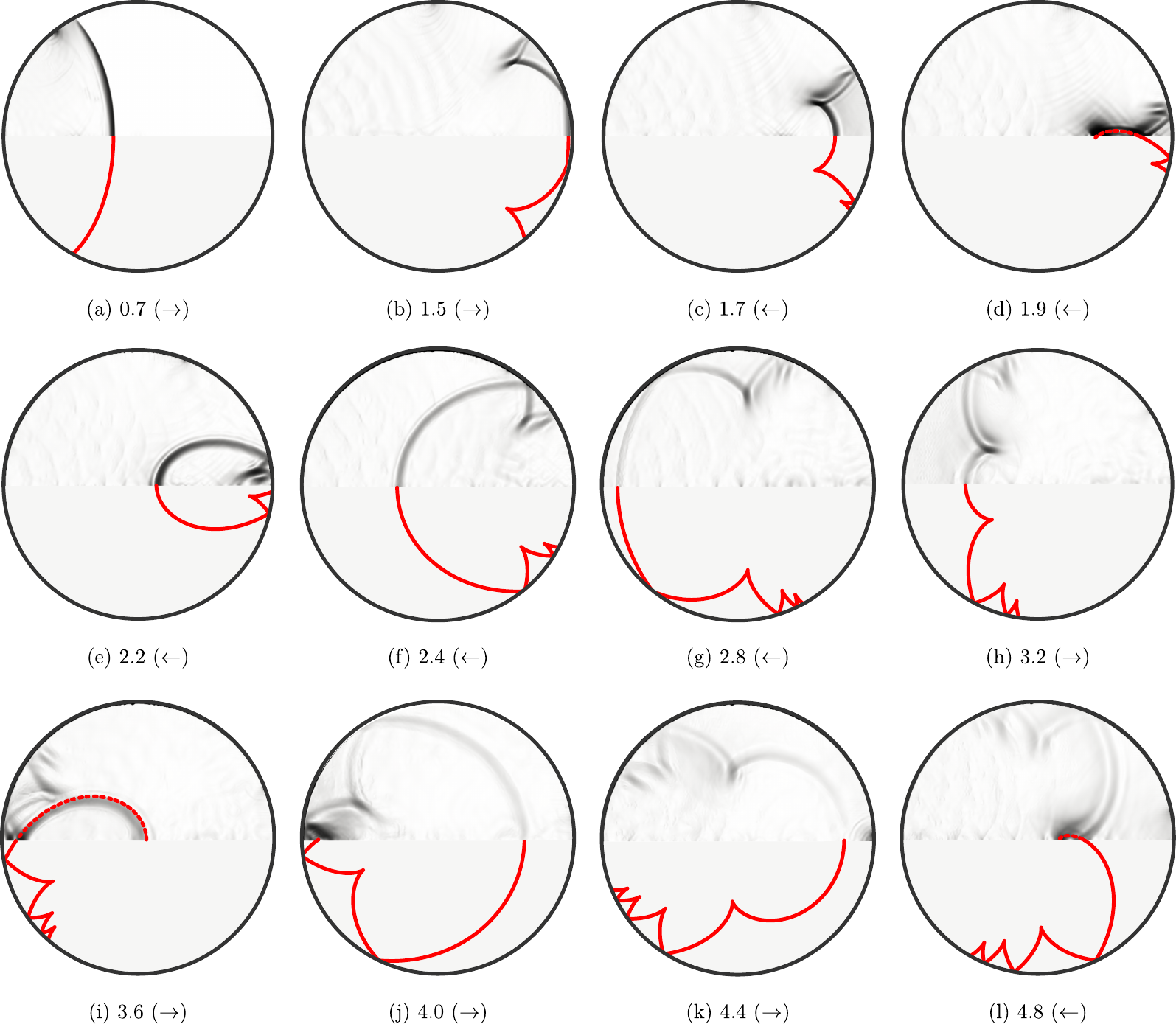}
	\caption{Comparison of the confined wavefront spatio-temporal dynamics theoretically predicted (lower half space) with numerical schlieren computed from simulations (upper half space). Theoretical wavefronts are given by Eq.~(\ref{eq:refracted_front}) for $n=2.12$ and displayed in red dotted line when they overlap the schlieren visualization. Times in $\SI{}{\micro\second}$ are given in the sub-captions and the arrow between parentheses indicates the direction of wave propagation.}
	\label{fig:Schlieren}
\end{figure*}

Figure~\ref{fig:Schlieren} displays the time evolution of the wavefront during a few round-trips. The upper half of each panel in the figure shows numerical Schlieren images\footnote{A non-linear scale has been applied to the images to enhance the visualization of the wave pattern\cite{quirk_dynamics_1996,meng_numerical_2015,johnsen_numerical_2009}} (magnitude of density gradient) from the simulations, whereas the lower half displays the corresponding fronts as predicted by ray theory, see Eqs.~(\ref{eq:refracted_front_x})--(\ref{eq:refracted_front_y}). The sequence runs from left to right and then top to bottom, and is not uniformly spaced in time (see caption for details). The time steps are selected so as to exhibit the principal features of the front evolution. It is clear from the comparison offered here that, with regard to both the shape and location of the wavefront, we have a remarkable agreement between the theoretical and numerical results. We point out that such an excellent agreement is obtained with no adjustable parameters.

\begin{figure}[h]
	\centering
	\includegraphics[width=300pt]{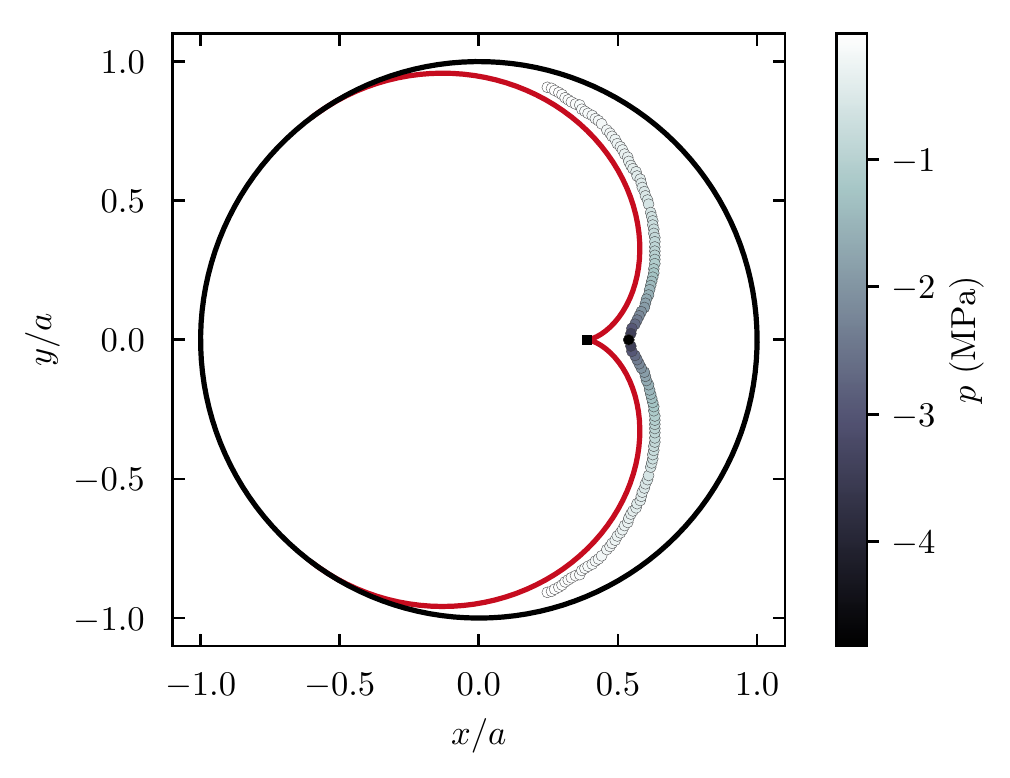}
	\put(-426,128){\makebox[0.5\textwidth][r]Caustic cusp~$\rightarrow$} 
    \put(-347,156){\makebox[0.5\textwidth][r]{\textcolor{NewRed}{Eqs.~(\ref{eq:caustic}a, b)~$\rightarrow$}}}
    \caption{The black solid line displays the droplet boundary ($n=2.12$). The red solid line is the caustic traced out by $\mathcal{F}_{2}$ wavefront cusp. Circle markers are extracted from numerical simulations and refer to the spatial location of the peak negative pressure over time, colored with the pressure magnitude.}
	\label{fig:figure7}
\end{figure}

In Fig.~\ref{fig:figure7}, we plot the caustic associated with the rays reflected once at the boundary of the droplet (red line in the figure), as expressed by Eqs. (\ref{eq:xcaustic})--(\ref{eq:ycaustic}). As we have already mentioned above, this surface is the locus of points where the ray intensity is the highest. Since upon reflection at the droplet interface the compressive wave is transformed into an expansion wave, it means that this caustic corresponds to the region of lowest pressure. The color-filled circles represent the lowest pressure as obtained from the simulation: the darker the color, the lower the pressure (see the color bar to the right of the figure). Each circle corresponds to the position of the front's cusp at different time instants, which are indicated by the same color code as for pressure, from white (\SI{1.07}{\micro\second}) to black (\SI{1.75}{\micro\second}). We note, in accordance with ray calculations, that the caustic's cusp is the point of lowest pressure. A slight shift between the caustic and the trajectory described by the front's cusp is observed. The shift is the largest on the $x$-axis. Equation~(\ref{eq:xcusp}) gives for $x_\mathrm{cusp} = 0.40 a$, which reasonably agrees with the simulation result of $0.51 a$. The time at which the caustic's cusp is reached is \SI{1.78}{\micro\second}, Eq.~(\ref{eq:tcusp}), which is very close to the value of \SI{1.75}{\micro\second} obtained from the simulation.

As already mentioned, our simulations do not take into account phase changes and interactions. This implies that we ignore the effects relevant to the dynamics of bubble formation and their feedbacks on the droplet evolution. Because these effects are expected to significantly alter the droplet dynamics, our simulation results can only be considered as valid up to the instant the first inception of cavitation is observed. Our concern here nevertheless is in determining incident shock conditions leading to cavitation inside a spherical droplet. For such a purpose, it seems reasonable to consider cavitation event is occurred in regions where pressure has dropped below some threshold value. 

\begin{figure*}
    \centering
    \begin{subfigure}[b]{.47\columnwidth}
        \includegraphics[width=.995\columnwidth]{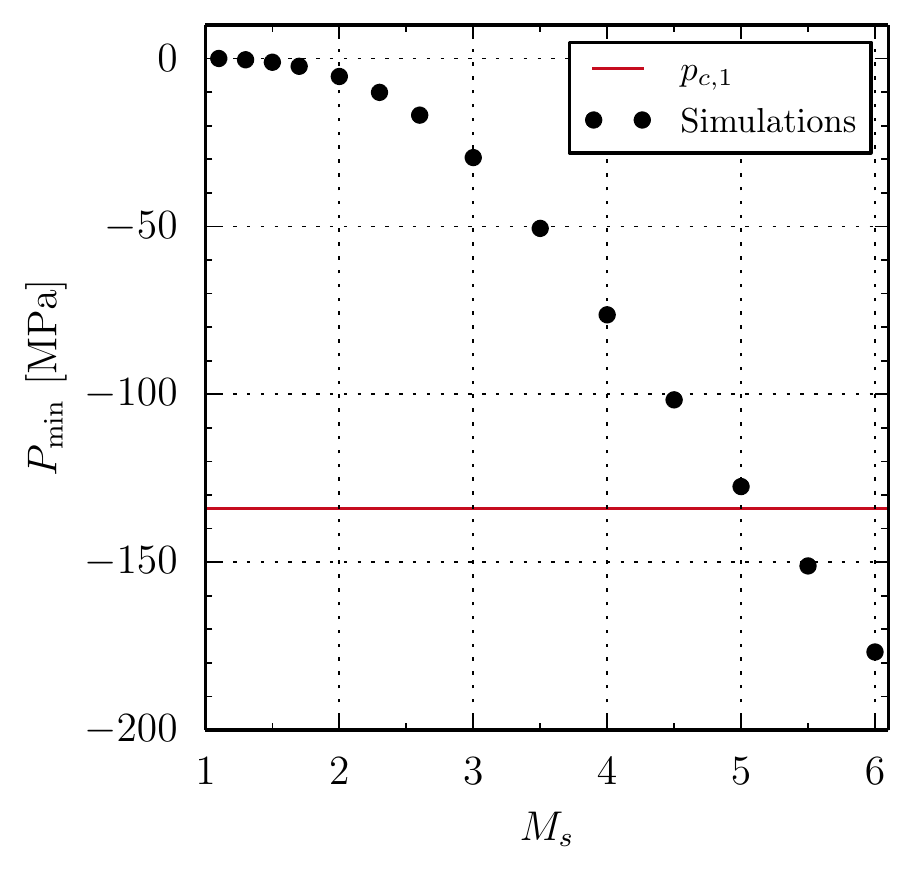}
        \caption{}
        \label{fig:MaPminFull}
    \end{subfigure}
    \begin{subfigure}[b]{.47\columnwidth}
        \includegraphics[width=.995\columnwidth]{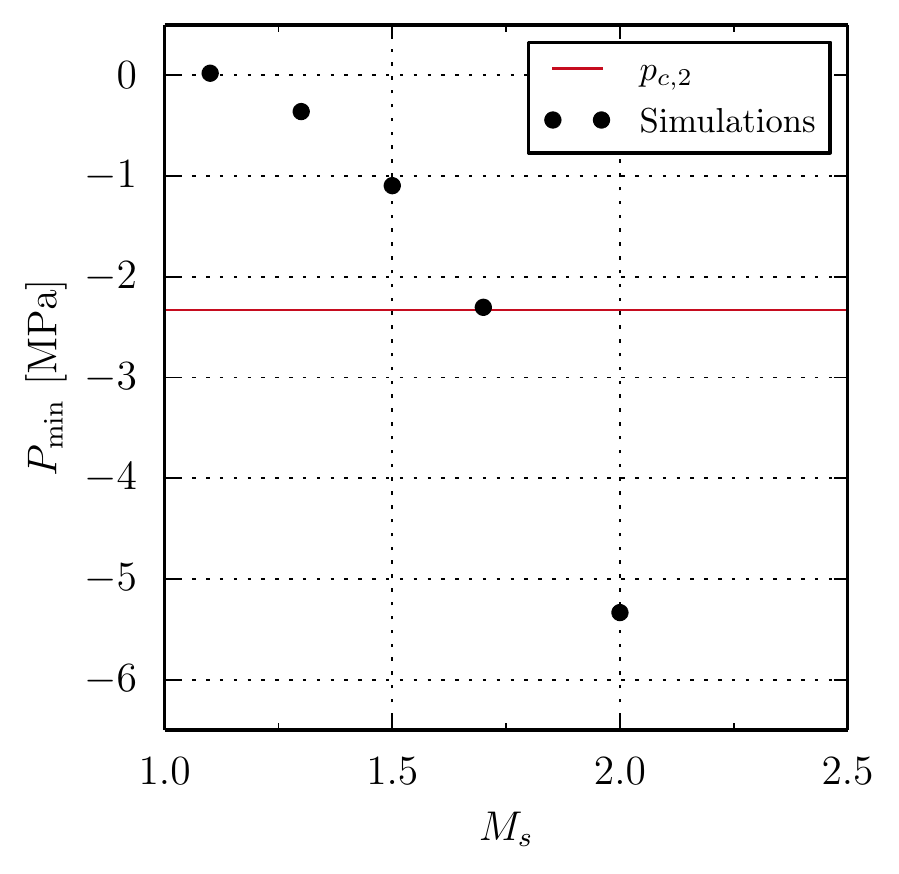}
        \caption{}
        \label{fig:MaPinZoom}
    \end{subfigure}
    \caption{Peak minimum pressure compared to (a) the cavitation pressure threshold $p_{c,1}$ as predicted by the classical nucleation theory and (b) the pressure thresholds $p_{c,2}$ experimentally reported by \citet{sembian_plane_2016}}
    \label{fig:MaPmin}
\end{figure*}

In order to identify conditions prone to the advent of cavitation zones, we have performed simulations for incident-shock Mach numbers, $M_s$, varying from 1.1 to 6.0. In Fig.~\ref{fig:MaPminFull}, we have plotted the lowest pressure reached inside the droplet, $P_\mathrm{min}$, downstream  of the first-reflected wave against $M_s$. The location of the minimum pressure corresponds approximately to that of the caustic's cusp. A few trends stand out from Fig.~\ref{fig:MaPminFull}. One is that $P_\mathrm{min}$ is negative for all $M_s$, with the exception at $M_{s}=1.1$ for which it is nearly zero. Additionally, $P_\mathrm{min}$ is a decreasing function of $M_s$, as should be expected. We note that $P_\mathrm{min}$ decreases slowly from 0 at $M_s=1.1$ to \SI{-5}{\mega\pascal} at $M_s=2.0$. On increasing $M_s$ beyond this latter point, $P_\mathrm{min}$ decreases at a higher pace. The most interesting aspect of this graph is the linear dependence of $P_\mathrm{min}$ on $M_s$, which is observed over the range $3.5\lesssim M_{s}\leq6.0$. At this stage, we cannot offer any explanation for this behavior. If we compare $P_\mathrm{min}$ with the pressure threshold $p_\mathrm{c, 1}$, whose location is displayed in Fig.~\ref{fig:MaPminFull} as the horizontal red line, we see that the regimes in which a liquid gas phase transition is likely to develop correspond to $M_s$ above 5. Figure~\ref{fig:MaPinZoom} is a zoom-in of Fig.~\ref{fig:MaPminFull} covering a much smaller $M_s$-range between 1.0 and 2.5. The horizontal red line indicates the location of $p_\mathrm{c, 2}$. In this case, we see that cavitation is likely to occur whenever $M_s$ exceeds 1.7. Such a critical value is almost three times smaller than the one obtained for pure water. As a final word, it can be mentioned that, in their study of shock-water column interaction, \citet{sembian_plane_2016} found that cavitation may arise for an incident shock wave Mach number greater than 2.4. This value is 50\% higher than what we found. The difference with our value of 1.7 can obviously be attributed to the higher degree of the rarefaction wave focusing achieved in a spherical droplet. In Table~\ref{tab:table1}, we report the shock wave Mach numbers for which homogeneous and heterogeneous cavitation is likely to occur according to our simulations.
\begin{table}
\caption{\label{tab:table1} Shock wave Mach numbers for which homogeneous and heterogeneous cavitation is likely to occur.}
\begin{ruledtabular}
\begin{tabular}{ccc}
      & \multicolumn{2}{c}{Shock wave Mach number, $M_{s}$} \\[3pt]
      \cline{2-3} \\
      Cavitation threshold & \citet{sembian_plane_2016} & Present simulations \\[3pt]
      $p_{c,1}$ & --    & $\approx5.0$\\
      $p_{c,2}$ & 2.4   & $\phantom{\approx}\,1.7$\\
\end{tabular}
\end{ruledtabular}
\end{table}

\section{Concluding remarks}
In this paper, we have examined the initial phases of the interaction between a planar shock wave in air and a spherical water-droplet. The analysis was conducted using ray theory, which provides analytical results that were compared with and complemented by numerical simulations. There are several remarks and conclusions that we consider of particular relevance concerning the results reported herein. First, we saw that the wavefront inside the droplet assumes a complex time-dependent shape whose dominant feature is the existence of cusp singularities. From ray calculations, it was clearly shown that these singular points result from the focusing process. Second, we derived parametric equations for the surface of confined wavefront. Comparisons with simulations showed that the front shape and its evolution are perfectly well described by these equations. This result offers a simple description of the geometry and the process of focusing of the wavefront during the interaction. Third, it was proved that each wavefront cusp traces out a surface, which is the caustic of the associated system of rays. The energy flux density turns out to be singular over these caustic surfaces. Although physically unrealistic, this singular behaviour may be interpreted as revealing regions of highest ray density. Furthermore, we showed that caustics exhibit cusps where the concentration of rays forming the caustics is the strongest. Finally, as a consequence of the wave impedance, the compression wave inside the water droplet reflects at the interface as an expansion wave, thereby forming low-pressure regions in the internal flow field. On the basis of cavitation pressure thresholds from the literature, we obtained the incident-shock-strength conditions under which a planar shock wave can cause cavitation within a droplet.

\begin{acknowledgments}
The authors gratefully acknowledge fruitful discussions with Tim Colonius from California Institute of Technology. This work was partially supported by the R\'{e}gion Nouvelle-Aquitaine as part of the SEIGLE project (grant number 2017-1R50115). The fisrt author acknowledges the support received by an ETH Zurich Postdoctoral Fellowship.
\end{acknowledgments}

\bibliography{biasiori2021}

\end{document}